# Active focal-plane coronagraphy with liquid-crystal spatial-light modulators: Broadband contrast performance in the visible

JONAS G. KÜHN,[1,*] AND POLYCHRONIS PATAPIS[2]

[1]*Physics Institute, University of Bern, Sidlerstrasse 5, 3012 Bern, SWITZERLAND*
[2]*Institute for Particle Physics and Astrophysics, ETH Zurich, 8093 Zurich, SWITZERLAND*
**jonas.kuehn@unibe.ch*

**Abstract:** The technological progress in spatial-light modulators (SLM) technology has made it possible to use those devices as programmable active focal-plane phase coronagraphic masks, opening the door to novel versatile and adaptive high-contrast imaging observation strategies. However, the scalar nature of the SLM-induced phase response is a potential hurdle when applying the approach to wideband light, as is typical in astronomical imaging. For the first time, we present laboratory results with broadband light (up to ~12% bandwidth) for two commercially-available SLM devices used as active focal-plane phase masks in the visible regime (640 nm). It is shown that under ideal or realistic telescope aperture conditions, the contrast performance is negligibly affected in this bandwidth regime, reaching sufficient level for ground-based high-contrast imaging, which is typically dominated by atmospheric residuals.

## 1. Introduction

Since the discovery of 51 Peg b [1], newly-discovered exoplanets in our galactic neighborhood have amounted to a few thousands confirmed detections, essentially through the use of radial velocity (RV, see e.g. [2]) and transit (see e.g. [3]) techniques. In parallel, observing exoplanets with direct "high-contrast" imaging (HCI) [4,5] carries a tremendous and synergistic potential, as HCI allows for in-situ observing of exoplanets in their host stellar environment, and opens the door to spectroscopic characterization of the atmosphere composition of exoplanets [6-9]. Yet, since the initial promising discoveries of Fomalhaut b [4] and HR8799 b,c,d,e [5,10] less than 10 years ago, the number of new confirmed detections of planetary mass objects through HCI has not exceeded a dozen, owing for a still large technological gap that needs to be addressed by the community.

Current-generation "planet-imager" instruments include the Gemini Planet Imager (GPI; see [11]), the Spectro-Polarimetric High-contrast Exoplanet Research (SPHERE) instrument [12] on the Very Large Telescope (VLT), and the Subaru Extreme-Adaptive Optics (SCExAO) facility [13,14]. These near-infrared (NIR) HCI instruments take advantage of the latest Adaptive Optics (AO) systems, using deformable mirrors (DMs) with several thousand actuators and running at 1 kHz or beyond, to correct for the atmospheric turbulences and reach the extreme AO (XAO) regime with a Strehl ratio (SR) in excess of 0.8. To this purpose, these state-of-the-art facilities make use of the latest wavefront sensor (WFS) techniques to correct for non-common path aberrations (NCPAs) present in the beam train between the AO WFS and the science focal-plane array (FPA) [15-17]. Then, optical coronagraphs are employed to mask the bulk of the stellar point-spread function (PSF), in order to reveal faint off-axis astrophysical sources. In practice, those coronagraphs can either block, or diffract away, the unwanted starlight in an intermediate focal-plane, using amplitude [18] or phase [19,20] patterns, or be located in an intermediate pupil-plane where they will reshape the stellar PSF to dig high-contrast regions in the field-of-view [21,22].

Still, in spite of these headways, the actual number of directly-imaged confirmed exoplanet detections has admittedly remained very modest so far [23-25]. Detecting exoplanets with HCI from the ground is admittedly a daunting task, not the least because of the extreme contrast ratio ($10^{-6}$ to $10^{-12}$) required at angular separations as small as few diffraction beam widths ($\lambda/D$ units). The upcoming class of extremely large telescopes (ELTs), bearing 25- to 40-m diameter primary mirrors, will ease some of these challenges, notably by improving sensitivity and angular resolution, but several of the aforementioned issues affecting direct imaging and coronagraphy will remain, or may even worsen. Those include residual differential atmospheric refraction, resolved nearby giant stars (a side-effect of using bigger telescopes), NCPAs, pupil registration stability, and non-ideal segmented telescope apertures. Part of these aspects are inherent to the larger ELTs apertures and support structures, but notably also arise from the segmented geometry of the ELTs primary mirrors, whose merit functions may even evolve over time due to dead, defective, altered (e.g. uneven reflectivity), or missing (e.g. during servicing) individual mirror segments. As for space-based observatories, the post-Hubble era is also seeing the appearance of segmented primary mirrors (starting with JWST, but also e.g. the 15-m LUVOIR concept [26] or the 6m-class flagship IR/O/UV space telescope recommended by the Astro2020 decadal survey [27]), with possibly similar issues, with even less margin for in-situ intervention. In the scenarios described above, it would be highly valuable if the high-contrast imaging instruments, including the coronagraph, were flexible enough to be remotely re-configured and re-optimized on-demand, with as little mechanical motion involved as possible.

In this context, we developed a research effort in the field of SLM-based active focal-plane phase coronagraphy, using off-the-shelf pixelated SLM display panels as active and adaptive focal-plane phase mask coronagraphs. Several exploratory research topics using SLMs for active coronagraphy were investigated, among them HCI of binary stars compatible with angular differential imaging (ADI) [28], implementation of complex pupil-conjugated focal-plane phase masks [29], self-calibrating coronagraphy by phase-shifting Zernike WFS of NPCAs [30], an illustration of the added potential of the technique for prototyping applications. All these capabilities are encompassed in an easy-to-use (software-only) HCI instrument concept, potentially allowing for optimal use of observing time in function of weather conditions (dynamical adaptation of the coronagraph pattern). More recently, this has led the SLM-based PLACID instrument concept to be retained for the new Turkish DAG telescope [31]. However, one should also stress that usage scenarios for SLMs in astronomical instrumentation have now definitively expanded beyond coronagraphy applications. Nowadays, liquid crystal on silicon (LCOS) SLMs are frequently used as active phase screens to introduce atmospheric turbulences for AO testbeds [32,33], even relying on real-data from AO telemetry of existing instruments, but also for speckles nulling over very large portions of the field-of-view [34].

Although LCOS SLMs (see e.g. [35,36]) provides exquisite spatial sampling (>90% fill factor, <10 um pixel pitch, several millions of pixels) enabling focal-plane applications, there are however a few numbers of potential shortcomings to deal with which might limit the usage scenarios for astronomical applications. First, those panels require linearly polarized light as an input. Hence at least 50% of the incoming unpolarized starlight has to be either thrown away, or redirected towards another optical arm, which put strong constrains to photon-starved applications in terms of sensitivity. Second, the SLM-induced phase shift is highly chromatic, as it relies on a scalar phase delay instead of a geometric phase shift like, e.g., the vector vortex [20] or vAPP coronagraphs [37]. Finally, the L-band (3.5 um) corresponds to a strong absorption band of LC material, in principle disqualifying this technology for mid-IR instrumentation. As the first of those issue is currently being actively addressed by SLM manufacturers developing prototypes of polarization-insensitive SLMs (PI-SLMs, based e.g. on [38]), in this work we will address the second issue listed above, namely the concern about the chromaticity of SLM phase retardance, and its impact on contrast performance in the context of HCI applications. Indeed, most direct imaging observations are nowadays conducted at NIR wavelengths, notably at J- (1.2 μm), H- (1.65 μm) and K-bands (2.2 μm), which corresponds to favourable atmospheric transmission windows of water vapor. However, these bandpass filters are usually at least 15 to 20% wide to exploit the atmospheric transmission efficiency as much as possible, hence the crucial need to assess SLM coronagraphic contrast capabilities beyond the monochromatic regime.

In practice, we will hereby explore moderate broadband regimes up to 12% bandwidth at 640 nm, within the sensitivity range of CCD detectors. We reckon that such a bandwidth is still at the low end of usefulness for science, as typical near-infrared astronomical filters are in the 15-20% bandwidth range (H- and Ks-bands), and the presented results should therefore be interpreted as a stepping stone towards SLM performance validation in conditions relevant to astronomical NIR imaging, a topic we are currently actively investigating.

## 2. Methods

### 2.1 Optical layout

In order to optically investigate the HCI contrast performance of SLM-based active focal-plane coronagraphy concepts in realistic conditions, we developed, and refined over the years, a dedicated testbed called "SWATCHi". SWATCHi stands for the Swiss Wideband Active Testbed for Coronagraphic High-contrast imaging, which was initially build at ETH Zurich, Switzerland, before being recently relocated at the Space Sciences Institute of the University of Bern, Switzerland. The SWATCHi optical setup (Fig.1) is an all-reflective (protected silver coated mirrors) high-contrast imaging testbed designed to test the coronagraphic contrast performance of various LCOS SLM panels that are operating as reflective focal-plane phase masks (FPMs), i.e. placed to operate in reflection in an intermediate focal-plane, under broadband light conditions. The SWATCHi bench is currently operating in the visible light regime using a low-noise 12-bits CCD camera (pco.pixelfly), but it is also in the process of being upgraded with a C-RED 3 InGaAs detector from First Light Imaging for NIR (J- to K-band) operations. Indeed, the setup operates with a versatile super-continuum light source (NKT Photonics SuperK Compact, 450-2400 nm spectral range) coupled to an intermediate stage fielding a 6-slots bandpass filter wheel, with the outgoing light being again fiber-fed up to the SWATCHi entrance focal-plane (see Fig.1 inset). For monochromatic light measurements, a pigtailed laser diode module (Thorlabs S1FC635, $\lambda$ = 635 nm) is used. An intermediate entrance pupil-plane (Fig.1) enables to insert various pupil masks (0.1-mm thick molybden laser cut disks, as illustrated on Fig.2) to simulate various telescope apertures at various focal-ratios (at F/30 or slower on the SLM). As shown on Figure 1, the beam f-ratio is then slowed down by a factor 1.5x before reaching the coronagraphic reflective focal-plane, where the LCOS SLM panel is inserted. A wire grid polarizer (Thorlabs WP25M-VIS) is placed in front to ensure only linearly-polarized light aligned along the extraordinary axis of the SLM LC molecules reaches the panel, while being installed in double-path configuration to mitigate potential internal depolarization effects. The geometric beam configuration in this plane plays a crucial role for the phase modulation: depending of the SLM pixel pitch (usually < 10 μm), the f-ratio has to be adjusted to ensure Nyquist or better spatial sampling of the PSF: as a general super-Nyquist rule, we settled on a sampling of 10 pixels per $\lambda/D$, in order to be able to implement complex FPM patterns while picking a value that is independent from the choice of pattern. In addition, the SLM is setup with an off-axis reflection angle of less than 5 deg to minimize inter-pixel crosstalk, according to the manufacturers' recommendations. An additional Lyot pupil-plane is then located downstream, to be able to insert Lyot masks (same manufacturing technique on Fig.2), before reaching the scientific focal-plane. Additionally, the last fold mirror is installed on a removable magnetic plate, enabling to mechanically switch to a pupil-imaging lens configuration, always useful for coronagraphic alignment and diagnostics (see Fig.3).

## 2.2 SLM devices and operations

The two commercially-available LCOS SLMs that were tested in this work are the PLUTO-014 panel from Holoeye GmbH (Germany) and the 1920x1080 panel from Meadowlark Co. (USA), corresponding to the two main SLM manufacturers in the Western hemisphere. Both display panels are standard "workhorse models" operating at video-rate (60 Hz), and are designed and calibrated for the visible light (633 nm) wavelength regime, but with direct equivalent for NIR wavelengths (1.55 μm) available off-the-shelf; this selection should facilitate direct comparison between the VIS and NIR regimes in the near future. Table 1 list most relevant parameters for each device, including pixel resolution, fill factor, diffraction efficiency and rms phase jitter. As depicted on Fig.1, both SLMs will be setup in an off-axis reflection configuration minimizing pixel crosstalk (~5 deg beam incident angle). This deliberate choice is to be compared with another classical alternative, which is the diffractive configuration, where the SLM is placed face-on to the beam, and a diffractive phase pattern is added to the programmed phase map in order to send the reflected light in the +1 or -1 diffraction order. Although the latter is potentially attractive for HCI (unwanted on-axis starlight has much less risk of contaminating the high-contrast channel), it would further complicate the matter when dealing with broadband light: the diffracted beam would translate wavelength dependency into spatial displacement, leading to a chromatic blur. A compensation mechanism (e.g. with a diffraction grating) could be devised, but the additional complexity would undoubtedly take a toll on the already modest optical throughput. Although our off-axis approach is in principle more straightforward, it will inevitably be sensitive to the on-axis "zero-order" leakage component, dominated by internal parasitic back-reflections inside the SLM panel. Albeit the SLM front glass plate is anti-reflection (A/R) coated (see Table 1), it is indeed not possible for the manufacturers to fully optimize the A/R coating at the glass-LC interface, as the refraction index of the latter varies with the applied voltage by essence. This effect is therefore expected to strongly affect the contrast results presented in this work.

Another important aspect of operating these SLM for imaging is to integrate frames long enough on the detector to sample out their 60 Hz video refresh rate (Table 1): typically, here we will use shutter times > 30 ms to ensure that we are not sensitive to refresh jitter effects. Finally, as previously mentioned, linearly polarized light has to be used when using the SLM for phase modulation. This polarization indirection has to be matched with the SLM LC extraordinary index direction, which interestingly is not the same for both manufacturers (Table 1). To optimize optical throughput, the liner polarizer (LP) direction is kept vertical (p-polarized) on the bench, and the SLM to be tested is rotated 90 deg accordingly. Although only very slow (F/143) beams well within the LP acceptance angle (+/- 15 deg) will be used for the presented experiments (Table 2), the converging nature of the beam incident on the SLMs (see Fig.1) could slightly affect its degree of liner polarization beyond the level of 100:1 that we are able to assess, hence potentially introducing extra leakage sources. To mitigate this, the linear polarizer is installed in a double-pass configuration along the diverging, respectively converging beam, in front (~50 mm) of the SLM panel, in order to ensure that there is no intermediate optics in between both components (Fig.1). We indeed observed a slight reduction in on-axis contrast performance at the few $10^{-3}$ level when using a single-pass LP configuration, a phenomenon that may be partially caused by the chromatic extinction ratio leakage from the LP (in the order of $10^{-3}$), but also possibly to depolarization effects inside the SLM 3-D structure.

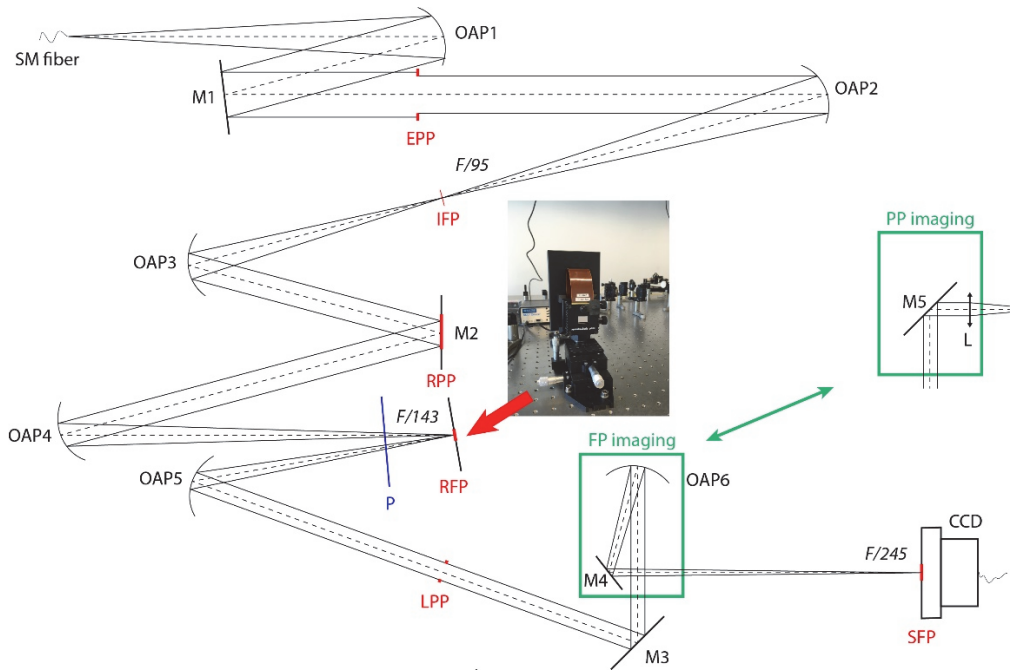

Fig. 1. Optical layout of the Swiss Wideband Active Testbed for Coronagraphy and High-contrast imaging (SWATCHi) testbed, dedicated to test LCOS SLM display panels as active focal-plane phase mask coronagraphs in broadband light in the VIS and NIR bands. OAP: off-axis parabolic mirror, M: mirror, P: wire grid polarizer, EPP: entrance pupil-plane, RPP: reflective intermediate pupil-plane, RFP: reflective (coronagraphic) intermediate focal-plane, with SLM (off-axis angle ~5 deg), LPP: Lyot pupil-plane, SFP: scientific focal-plane. The green insets denote switchable focal- (FP) or pupil-plane (PP) imaging stages.

Table 1. Main specifications of the two SLMs investigated in this work.

|  | SLM 1 | SLM 2 |
|---|---|---|
| Manufacturer | Holoeye GmbH | Meadowlark Corp. |
| Model # | PLUTO-VIS-014 | P1920-0635-HDMI |
| Resolution | 1920 x 1080 px | 1920 x 1152 px |
| Calibration wavelength | 633 nm | 633 nm |
| Pixel pitch | 8 um | 9.2 um |
| Fill factor | 93 % | 95.7 % |
| Response time | 16 ms | 12 ms |
| Diffraction efficiency | ~ 86% | ~ 78% |
| Phase jitter | < 22 nm rms | < 12 nm rms |
| Front A/R coating reflectivity | ~ 0.5 % | < 1% |

*2.3 Choice of filters and aperture masks*

Three different operating wavelength regimes are investigated here: (i) monochromatic light (633 nm laser diode); (ii) 6% wide broadband light (supercontinuum with a bandpass filter centered at 640 nm); (iii) and 12% wide broadband light (same as (ii) with a wider bandpass filter). The specifics of each light source and bandpass filter configuration are summarized in Table 2. Given that the SWATCHi testbed is fed with SM fibers as inputs (Fig.1), the change of configuration is straightforward and entirely transparent in terms of optical alignment.

The main design criterion when selecting the aperture masks is the diameter of the primary aperture, which sets the focal ratio for the entire beam train. Choosing a slow enough focal ratio is critical for SLM focal-plane applications, as the PSF has to be sufficiently well sampled to be able to program non-trivial phase patterns. As mentioned above, and given that the direct imaging applications we are interested in here do not require large on-sky field of views, we opted for the Nyquist "super-sampling" criterion, hence requiring a spatial sampling of the PSF of 10 SLM pixel per $\lambda/D$ units in the coronagraphic focal-plane (CFP in Fig.1). It is of course possible, in principle, to increase the spatial sampling further, but this would limit the on-sky field-of-view in the NIR wavelength range to less than 10 x 10 arcsec on 4- to 5-m telescopes, seriously restricting the discovery space in the longer run. Knowing the largest pixel pitch of 9.2 um for the Meadowlark unit (see Table 1), the sampling of 10 pixels per $\lambda/D$ units sets the fastest acceptable focal-ratio to achieve the desired PSF sampling. In practice, we opted for 4-mm apertures radii for the entrance pupil-plane masks, yielding a f-ratio of F/143 at the SLM location, corresponding to a ~91 um spot size for the wavelength regimes considered. For the centrally-obscured apertures, mimicking a real-life telescope pupil with the secondary mirror obstruction, we went for an obscuration ratio of ~0.3, along the lines of the 5-m Palomar Hale or 8-m Subaru Telescope configurations. This is however admittedly not capturing the scenario for a VLT-like pupil, which has a considerably more modest obscuration ratio in the order of 0.15. Finally, the support struts (spiders) are obviously also manufactured into the pupil masks using laser cutting (see Fig.2), but manufacturing limitations results into about 10x larger relative thickness as compared to real-world values. For the two above reasons, the centrally-obscured configuration presented here should be seen as worst-case scenarios, or upper contrast limits, for the purpose of evaluating SLM coronagraphic performance with obstructed apertures.

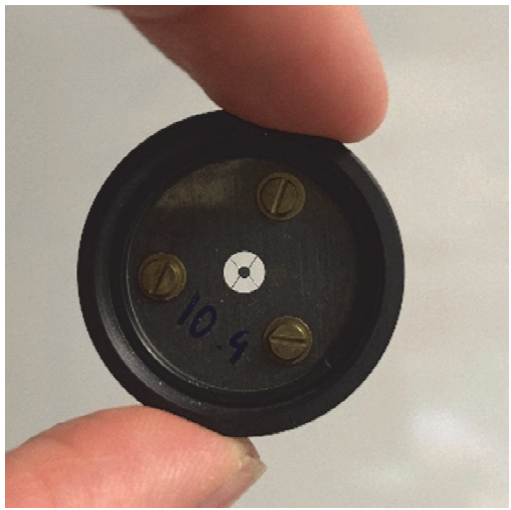

Fig. 2. Example of laser-cut pupil masks and Lyot stops (0.1-mm thick molybden on 25.4 mm mounts).

Table 2. Bandpass filter and aperture mask configurations.

| Bandpass Filter | Narrow-band | Broadband |
|---|---|---|
| Manufacturer | Semrock | |
| Filter model # | SEMFF01-640/40-25 | SEMFF02-641/75-25 |
| Central wavelength | 640 nm | 641 nm |
| Bandwidth [nm] | 40 nm | 75 nm |
| Bandwidth % | 6 | 12 |
| Entrance Aperture Mask | Unobscured Pupil | Centrally-obscured Pupil |
| Outer (primary) diameter $D_p$ | | 4 mm |
| Corresponding focal-ratio on SLM | | F/143 |
| Secondary diameter $D_s$ | N/A | 1.3 mm ($D_s/D_p \sim 0.3$) |
| Support struts thickness $t_s$ | N/A | 0.05 mm ($t_s/D_p \sim 0.0125$) |

## 3. Results

### 3.1 Ideal unobstructed apertures

We hereby present contrast measurement results obtained using the SWATCHi tesbed of Fig.1 and the two SLMs of Table 1 programmed with coronagraphic phase helix patterns with a central discontinuity (vortex of charge n=2 or n=4, i.e. the number of phase jumps per revolution), classically used for high-angular resolution HCI (inner-working angle [IWA] down to 1.2 $\lambda$/D for n=2 vortex, resp. 1.8 $\lambda$/D for n=4 vortex) [14,20]. The only equivalent phase pattern in terms of popularity in the HCI community is the four-quadrant phase mask (FQPM) [19]. But the geometry of the latter does not require the high spatial sampling of an SLM, and its IWA, defined as the half-transmission point of the coronagraph in function of angular separation from the host star, is at least three times worse when compared to an n=2 vortex. Three different operating wavelength regimes are then compared, namely monochromatic ($\lambda$ = 633 nm) using a pigtailed laser source module, and two configurations using the supercontinuum light source combined with the filter sets of Table 2: 6% (aka "narrow-band") and 12% (aka "broadband") bandwidth centered around $\lambda$ = 640 nm. Here we first focus on the "ideal scenario" for high-contrast imaging, i.e. an entrance pupil mask mimicking an "unobscured aperture" (e.g. an off-axis telescope), corresponding to an optimal Strehl ratio and Airy pattern. Overall, this corresponds to 18 different optical configurations – (2 coronagraphic + 1 for non-coronagraphic photometry), times 3 bandwidth regimes, times 2 SLM models – and each time we acquire sets of 100 focal-plane images with the pco.pixelfly CCD camera, and 20 pupil-plane images (10 with Lyot stop, 10 without), to mitigate readout noise, and improve SNR and dynamic range. Dark, background and flat calibration frames are also acquired for each integration time, typically ranging from 30 ms (minimum to avoid pixel flickering with the 60 Hz SLMs) to 2500 ms to avoid saturation, depending on the configuration. Bad pixel maps are also generated using the calibration frames.

Figure 3-Top presents typically obtained median-combined and photometry-adjusted processed final frames, in this case for SLM #1 at 12% bandwidth around 640nm. A gross estimate of the peak-to-peak attenuation can be derived from the ratio of integration times to avoid saturation for the coronagraphic (Fig.3-Top left column) and non-coronagraphic cases (Fig.3-Top middle and right columns), in this case about 43:1 (1300ms/30ms). Table 3 provides a summary of several relevant coronagraphic contrast metrics measured for all the vortex n=2 configurations investigated, and Figure 4-Top shows multiple azimuthally-averaged raw contrast curves for both SLMs, and the two extreme cases in terms of wavelength regime (monochromatic vs. 12% bandwidths). In particular, the first row of Table 3 illustrates how well each SLM device performs in ideal conditions (monochromatic regime), and it can be noted that the Holoeye panel slightly outperforms the Meadowlark display, likely thanks to an optimized A/R coating. We note that Figures 4 and 5, as well as Table 3, do not include results for the n=4 vortex configurations for the sake of clarity, as there were no discernable differences in contrast numbers observed (down to the few $10^{-4}$ contrast level).

### 3.2 Centrally-obstructed apertures (real telescope case)

An extra 18 similar optical configurations as in Section §3.1 were measured, but using a centrally-obscured entrance pupil mask (see Table 2), mimicking a more realistic contemporaneous ground-based telescope. As can be readily observed from the imaging examples of Fig. 3-Bottom, such a pupil reshapes (widens) the PSF with a stronger 1$^{st}$-order Airy ring, and introduces two leakage sources inside the high-contrast region of the the post-coronagraphic Lyot pupil plane: one from the spider support struts (usually negligible on-sky), and one from the secondary mirror obstruction, particularly for the less aggressive n=2 vortex pattern (an n=4 vortex has twice more phase jumps per revolution, yielding better null depth and robustness to low-order aberrations, at the cost of degrading the IWA by about factor two). Both are ultimately masked by an oversized/undersize Lyot stop nevertheless (not discussed here, see e.g. [14] for typical values for sizing the Lyot stop), hence this does not affect contrast but rather throughput. Examples of retrieved contrast curves are plotted on Fig.4-Bottom, and all the measured contrast metrics are presented in Table 3. To allow for direct comparison with a "typical" on-sky coronagraphic setup downstream on an exAO-equipped 8-m telescope, a raw contrast curve of Procyon observed with the Subaru Telescope's SCExAO n=2 vector vortex coronagraph at H-band (see Fig.5 in [14]) is also plotted on Figure 4-Bottom. One can notably observe that even the modest SLM contrast performance presented there can exceed achievable on-sky raw contrast on ground-based observatories equipped with state-of-the-art equipment.

### 3.3 Contrast vs. bandwidth

Of particular importance for real-world astronomical imaging applications is the dependence of the various contrast metrics, summarized on Table 3, to the optical wavelength regime (bandwidth). Figure 5 attempts to provide an oversight on this phenomena, by plotting the three contrast metrics of Table 3 in function of the optical bandwidth, for both SLMs and each entrance pupil configuration. As with contrast curves of Fig.4, little contrast variation vs. bandwidth is observed, albeit slightly for SLM #1 on-axis rejection, but the degradation in contrast when dealing with the centrally-obscured aperture is clear, especially inside 2 $\lambda$/D.

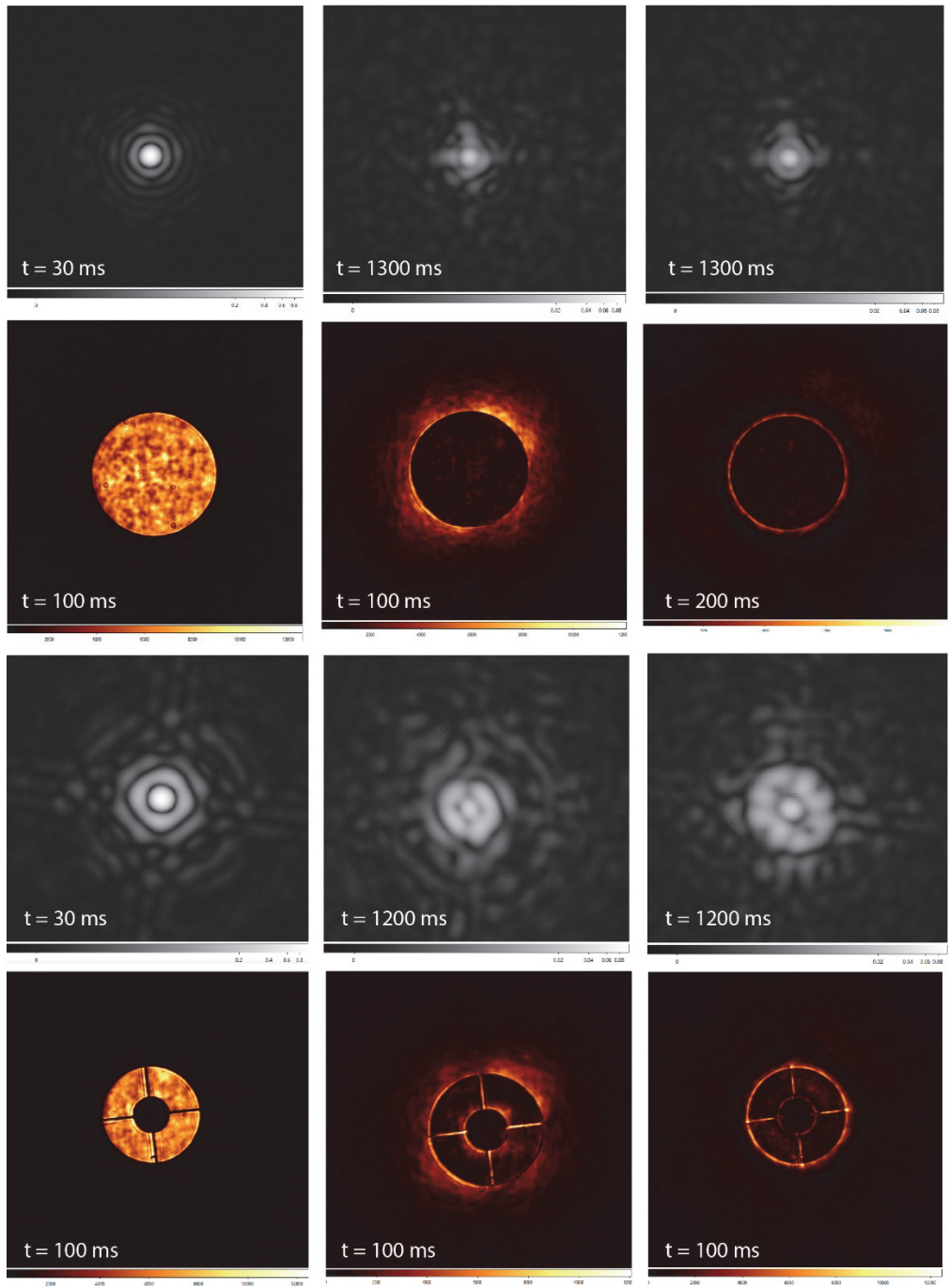

Fig. 3. Subset of typical post-processed median-combined PSF and pupil-plane datasets, with indicative camera integration times to appreciate the contrast gain.
(Left column) Non-coronagraphic datasets; (Center column) Coronagraphic (vortex n=2) datasets; (Right) Coronagraphic (vortex n=4) datasets. (Top half) Unobscured ideal entrance pupil case; (Bottom half) Centrally-obscured entrance pupil (see Table 2 for details).

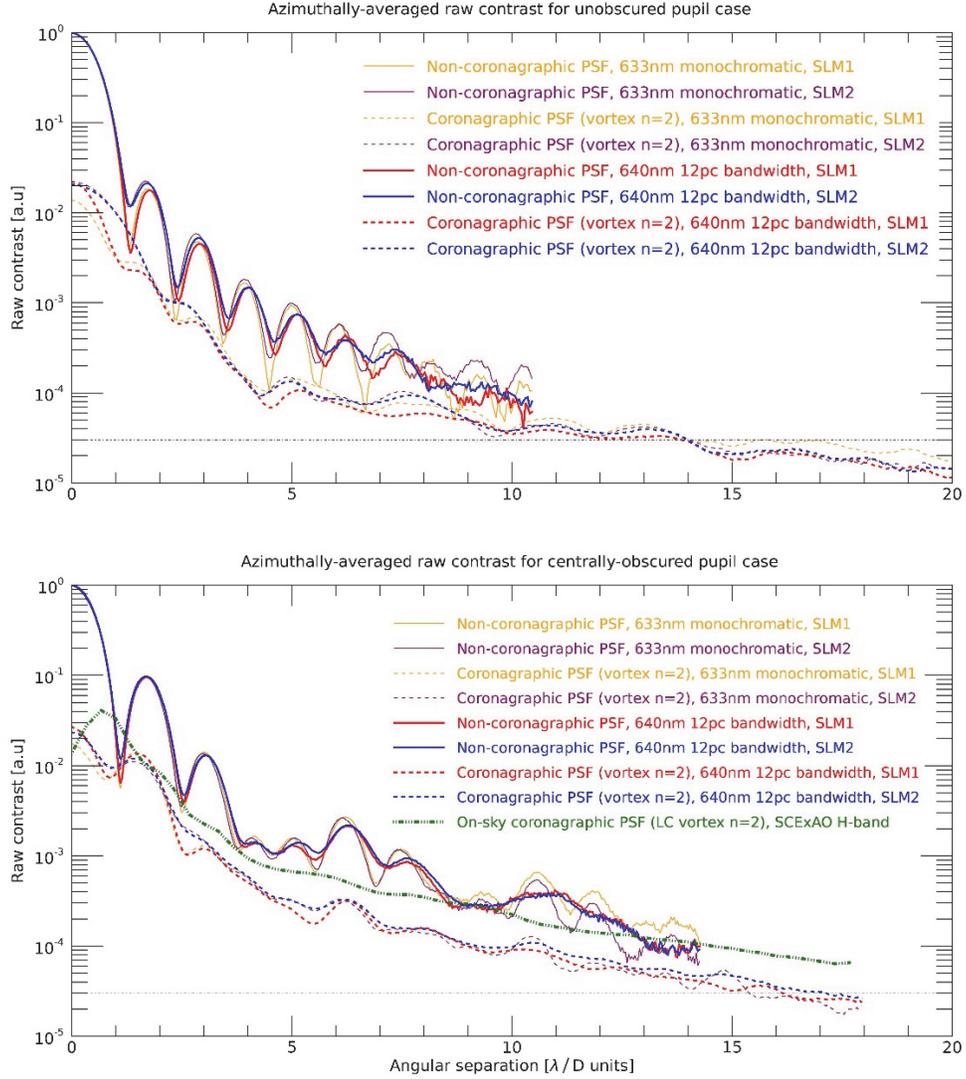

Fig. 4. Raw azimuthally averaged contrast curves for both SLMs with an (Top) unobscured ideal entrance pupil and a (Bottom) centrally-obscured entrance pupil (see Table 2 for details), where a typical on-sky raw contrast curve is also added for comparison (example of Procyon observed with SCExAO n=2 liquid crystal vector vortex coronagraph in the H-band, taken from Fig. 5 in [14]). The horizontal dashed line corresponds to the theoretical photon noise contrast limit in the non-coronagraphic case.

## 4. Discussion

Overall, when used as programmable focal-plane phase coronagraphs, the peak-to-peak stellar attenuation delivery by the two off-the-shelf LCOS SLMs that we tested is admittedly quite modest, in the range of 50:1 to 70:1 ($2 \cdot 10^{-2}$ to $1.5 \cdot 10^{-2}$). This is the case for the ideal case, that is with an unobstructed pupil in monochromatic light conditions. In this scenario, the on-axis null is probably dominated by the parasitic reflections internal to the SLMs, notably at the glass/air front interface (about 0.5-1% depending on the A/R coating and the SLM, see Table 1) and also the less known glass/LC interface. We can also observe a "butterfly" or "flower petals" geometry of the coronagraphic PSF residuals close to the star (see Fig.3, medium row) that nullifies any contrast gain around 1 λ/D angular separation, an effect likely caused by aliasing issues close to the center of vortex patterns. While the consequence of using a more realistic centrally-obscured pupil is non-negligible, plateauing the on-axis contrast to $2·10^{-2}$ and degrading the raw contrast at 2 λ/D by a factor ~4 (see Fig.5), the actual impact of broadening the light source spectra up to 12% bandwidth is less clear (see Fig.4 and 5). This is likely linked to far more negligible contribution of chromatic leakage to the overall contrast for this range of bandwidth. Indeed, using the approximation of [39] for the chromatic leakage for a 12% bandwidth centered at 640 nm, one typically gets a chromatic null $N_{chrom} \sim 2.8·10^{-3}$, roughly an order of magnitude less than the joint contribution of the imperfect A/R coatings internal to the SLMs, and the centrally-obscured non-ideal aperture. In conclusion, the former is clearly the main leakage contributor in the "zero-order" (non-diffractive) SLM geometric implementation that is used in all the experiments presented here.

Table 3. Measured contrast values for both SLMs, under various aperture and bandwidth configurations.

| Pupil geometry | | SLM 1 (Holoeye) | | SLM 2 (Meadowlark) | |
|---|---|---|---|---|---|
| | | Unobscured | Centrally-obscured | Unobscured | Centrally-obscured |
| Monochromatic | On-axis null | $1.4 \pm 0.05·10^{-2}$ | $1.8 \pm 0.1·10^{-2}$ | $2.2 \pm 0.05·10^{-2}$ | $2.4 \pm 0.1·10^{-2}$ |
| | Contrast @2λ/D | $1.3 \pm 1·10^{-3}$ | $6.3 \pm 3·10^{-3}$ | $1.3 \pm 1·10^{-3}$ | $5.9 \pm 3·10^{-3}$ |
| | Contrast @5λ/D | $1.4 \pm 1·10^{-4}$ | $3.8 \pm 3·10^{-4}$ | $1.5 \pm 1·10^{-4}$ | $3.2 \pm 3·10^{-4}$ |
| 6% bandwidth | On-axis null | $2.0 \pm 0.05·10^{-2}$ | $2.3 \pm 0.1·10^{-2}$ | $2.3 \pm 0.05·10^{-2}$ | $2.6 \pm 0.1·10^{-2}$ |
| | Contrast @2λ/D | $1.2 \pm 1·10^{-3}$ | $6.6 \pm 3·10^{-3}$ | $1.2 \pm 1·10^{-3}$ | $6.2 \pm 3·10^{-3}$ |
| | Contrast @5λ/D | $1.2 \pm 1·10^{-4}$ | $2.5 \pm 3·10^{-4}$ | $1.5 \pm 1·10^{-4}$ | $3.2 \pm 3·10^{-4}$ |
| 12% bandwidth | On-axis null | $2.2 \pm 0.05·10^{-2}$ | $2.7 \pm 0.1·10^{-2}$ | $2.1 \pm 0.05·10^{-2}$ | $2.3 \pm 0.1·10^{-2}$ |
| | Contrast @2λ/D | $1.2 \pm 1·10^{-3}$ | $6.9 \pm 3·10^{-3}$ | $1.2 \pm 1·10^{-3}$ | $6.5 \pm 3·10^{-3}$ |
| | Contrast @5λ/D | $1.0 \pm 1·10^{-4}$ | $2.5 \pm 3·10^{-4}$ | $1.3 \pm 1·10^{-4}$ | $3.2 \pm 3·10^{-4}$ |

The implications of these results are essentially two-folds: (i) albeit modest, the coronagraphic contrast performance of SLM-based focal-plane phase coronagraphs are sufficient for most ground-based instruments, using current-generation XAO systems (see e.g. a typical on-sky contrast curve on Fig.4), up to 10-15% chromatic bandwidths; and (ii) the achievable contrast levels are however clearly not good enough for HCI in absence of atmospheric residual aberrations, i.e. for space-based coronagraphs, as e.g. considered for the Roman Space Telescope [40]. It remains to be seen if such behavior holds true for slightly larger bandwidth regimes (20%, or even 25%), which would mean that SLM-based coronagraphs can be used with any of the most common astronomical imaging filters from the ground. In this regard, we are currently in the process of evaluating NIR coronagraphic performance of SLMs with an astronomical H-band filter (20% bandwidth at 1.6 μm), with a goal to reach up to Ks-band (2.2 μm) in the end. Indeed, whereas in principle the H-band is easily accessible with off-the-shelf SLMs designed for telecommunication band (1550 nm), Ks-band requires some in-house customization from the manufacturer to achieve sufficient retardance (i.e. one full wave of "stroke"). Furthermore, NIR SLMs' thermal properties will need to be investigated in this wavelength regime, notably to determine if thermal background light is compatible with high-contrast imaging.

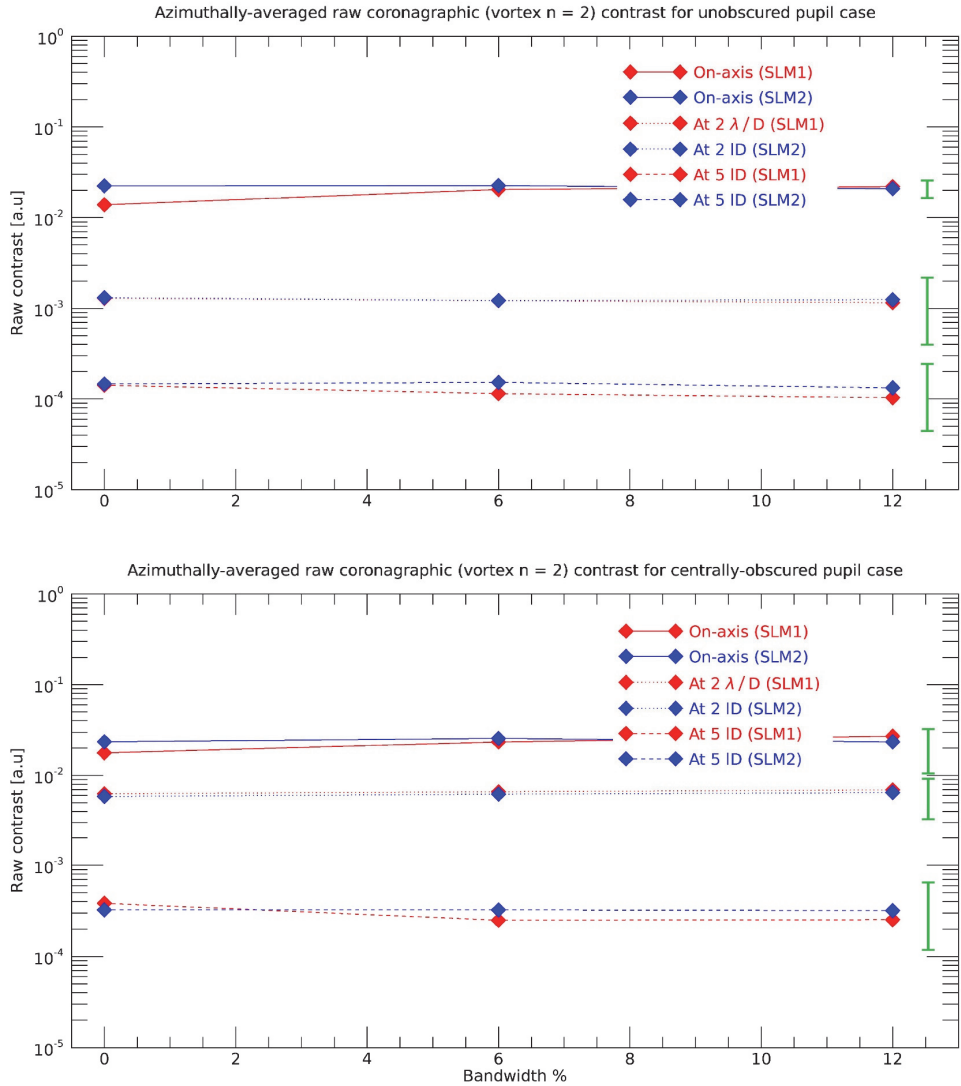

Fig. 5. Azimuthally-averaged raw contrast levels (on-axis, at 2 λ/D and at 5 λ/D) for both SLMs in function of bandwidth with an (Top) unobscured ideal entrance pupil and a (Bottom) centrally-obscured entrance pupil (see Table 2 for details). The green error bars depict the level of uncertainty (standard deviation in time for the on-axis null, in azimuth for off-axis contrasts) in each case.

In this regard, there are still several technological gaps and detailed studies that needs to be cleared in order to elevate the readiness level of SLM-based coronagraphs towards competitiveness for cutting-edge direct imaging science. An obvious disadvantage of SLMs is the requirement for linearly-polarized light, and so, one should carefully follow the ongoing development in this field, as prototypes of polarization-insensitive SLMs (PI-SLMs) are progressively introduced to the market. Among other approaches, we note that there have been recent attempts to prototype SLM panels with the quarter-waveplate (QWP) layer solution proposed decades ago in [38], and that prototypes of PI-SLMs are now available to order by companies such as Meadowlark, which we also intend to procure and investigate in the near future. As we tested here off-the-shelf SLM devices not particularly optimized for HCI applications, more in-depth device selection and testing should be undertaken in the future. Is high-phase resolution (up to 16 bits, vs. 8 bits here) or high-retardance resolution SLMs better suited for coronagraphy? How does this, and key SLM parameters like spatial sampling (# of pixels per $\lambda/D$ units), phase flicker and fill factor, influence on-axis and off-axis contrast, and is there an optimal design enabling the manufacturer to minimize SLM internal reflections? Or is there a viable off-axis diffractive scheme for using the SLMs in the context of HCI, preserving throughput and spatial resolution for imaging faint astrophysical sources? What is the full-Stokes response of those SLMs for coronagraphic use scenarios, and how accurately polarization needs to be controlled to reach even deeper contrast levels? All these questions deserve to be addressed in-tandem with SLM manufacturers, but also through simulations and more advance metrology schemes, like fast microscopic interferometric schemes to probe the phase retardance at the pixel scale in real-time. The problematic described above is particularly timely, as the range of usage scenarios for SLMs in astronomical imaging is quickly expanding, beyond the presented adaptive focal-plane coronagraphy applications. Still, even in the rather focused SLM usage scenario presented here, we note that a wealth of research avenues is now created to investigate more complex coronagraphic phase patterns that have been recently proposed, for example so-called wrapped vortices [41] that are optimized for broadband operations. In the meantime, high-speed SLMs (typ. 500-700 Hz), are being introduced to the market, which could expand the technology applicability to high-speed imaging with the newest kHz detectors (eAPDs, MKIDs etc.). This would enable to get in the range of the atmospheric coherence time for near-infrared regime, potentially opening a wealth of capabilities for wavefront control, lucky imaging and coherent differential imaging.

**Funding.** Swiss National Science Foundation (SNF Ambizione grant #PZ00P2_154800, NCCR PlanetS #TP_SF6).

**Acknowledgments.** The authors would like to thank Prof. H. M. Schmid, Dr. X. Lu and Mr. M. Arikan for their support and contribution to this research while at ETH Zurich, as well as Prof. N. Thomas and Prof. T. Feurer for providing laboratory space to the project at the University of Bern.

**Disclosures.** The authors declare no conflicts of interest.

**Data availability.** Data underlying the results presented in this paper are not publicly available at this time but may be obtained from the authors upon reasonable request.